\documentclass[usenatbib]{mnras}
\usepackage[varg]{newtxmath}
\usepackage[T1]{fontenc}
\usepackage{graphicx}
\usepackage[figuresleft]{rotating}
\usepackage{lscape}
\usepackage{journals}
\usepackage{comment}
\usepackage{multirow}
\usepackage{rotating}
\usepackage{xcolor}
\usepackage{textcomp}
\usepackage{ulem}

\definecolor{grey}{rgb}{0.4,0.6,0.6}
\definecolor{brown}{rgb}{0.65,0.16,0.16}
\definecolor{darkgreen}{rgb}{0.0,0.6,0.0}

\defcitealias{DiazGimenez+20}{Paper~I}
\defcitealias{DiazGimenez+21}{Paper~II}

\begin{document}

\title[Compact Groups in 3D]{Compact groups from semi-analytical models of galaxy formation -- III: purity and completeness of Hickson-like catalogues}
\author[Taverna et al.]
{A. Taverna$^{1,2}$\thanks{ataverna@unc.edu.ar},
E. D\'iaz-Gim\'enez$^{1,2}$,
A. Zandivarez$^{1,2}$,
G. A. Mamon$^{3}$ 
\\
\\
$1$ Universidad Nacional de C\'ordoba (UNC). Observatorio Astron\'omico de C\'ordoba (OAC). C\'ordoba, Argentina\\
$2$ CONICET. Instituto de Astronom\'ia Te\'orica y Experimental (IATE). C\'ordoba, Argentina\\
$3$ Institut d'Astrophysique de Paris (UMR 7095: CNRS \& Sorbonne Universit\'e), Paris, France}

\date{\today}
\pagerange{\pageref{firstpage}--\pageref{lastpage}}
\maketitle
\label{firstpage}

\begin{abstract}

Many catalogues of isolated compact groups of galaxies (CGs) have been extracted using Hickson's criteria to identify isolated, dense systems of galaxies, with at least three or four galaxies concordant in magnitude and redshift. But is not clear to what extent the catalogues of CGs are complete and reliable, relative to 3D truly isolated, dense groups.
Using five different semi-analytical models of galaxy formation (SAMs), we identify isolated dense groups in 3D real space, containing at least three galaxies. 
We then build mock redshift space galaxy catalogues and run a Hickson-like CG finder.
We find that the Hickson-like algorithm in redshift space is poor at recovering 3D CGs of at least 3 galaxies, with a purity of $\sim 10\%$ and a completeness of $\sim 22\%$. 
Among the $\sim 90\%$ of spurious systems, typically $60\%$ are dense structures that failed the 3D isolation criteria, while the remaining $40\%$ are chance alignments of galaxies along the line of sight, nearly all of which are within regular groups, with some variation with the SAM used for the analysis. In other words, while only 10\% of CGs are isolated dense groups, as intended, half are dense structures embedded within larger groups, and one-third are chance alignments within larger groups.
The low completeness of the extracted CG sample is mainly due to the flux limits of the selection criteria.
Our results suggest that a new observational algorithm to identify compact groups in redshift space is required to obtain dense isolated galaxy systems.

 \end{abstract}

\section{Introduction}
Compact groups (CGs) are highly-dense galaxy systems characterised by low membership and local isolation, creating an ideal scenario for dynamical interactions between their members.

In an attempt to select groups that satisfy these characteristics, \cite{Hickson82} performed a visual selection of compact groups guided by three general properties that these groups must fulfil: population, isolation, and compactness. 
Briefly, he systematically examined with a hand lens the red prints of the Palomar Observatory Sky Survey and selected groups with four or more galaxies within a range of three magnitudes from the brightest (population), without other bright galaxies in the surrounding ring of radius three times the angular radius of the smallest circumscribed circle defined by the galaxy members (isolation), and using a limiting group surface brightness as indicator of compactness. When redshift information was available for the galaxy members \citep{Hickson92}, a velocity concordance criterion of $1000 \rm km/s$ was included and groups of at least three concordant galaxies were accepted.  

To perform statistical studies of these peculiar systems, until today astronomers apply similar criteria to different observational samples of galaxies but using automatic algorithms \citep{Prandoni+94, Iovino+02, Iovino+03, Lee+04, deCarvalho+05, McConnachie+09, DiazGimenez+12, sohn+15, DiazGimenez+18, Zheng+20}. 
Other authors identified compact groups using percolation algorithms (for example  the friends-of-friends algorithm) but trying to mimic the main characteristics of the Hickson CGs \citep{Barton96, Allam+00, Focardi&Kelm02, hernandez-fernandez+15, sohn+16}. 
Despite the use of the same criteria, the automatic samples are not able to recover the original visual selection of compact groups. 
The automatic search of compact groups usually improves the completeness of the samples relative to the visual selection, but at the same time includes objects that were disregarded on purpose in the visual selection, leading to samples that are intrinsically different to the original Hickson Compact Groups.  

Due to the wide diversity of catalogues, the identification of CGs has not been homogeneous. The surveys have different apparent magnitude limits, different wavebands, and may or may not have spectroscopic information. Galaxies selected in different wavebands will form groups with different physical characteristics \citep{Taverna+16}. In addition, the limits imposed in the criteria are arbitrary and can change depending on the identifier.

Several studies have questioned the homogeneity of the samples identified by different authors.
\cite{DiazGimenez+12} identify 85 CGs in the Two Micron All Sky Survey (2MASS) \citep{2mass2} and, after comparing with the \citeauthor{Hickson82} sample (HCG), they found 16 CGs in common.
\cite{hernandez-fernandez+15} found that 20\% of their star-forming CGs identified in projection in FUV-band in the Galaxy Evolution Explorer (GALEX) survey have a counterpart in previous catalogues (only 8\% have a counterpart with redshift information).

Comparing different samples of CGs, even when the criteria are applied to the same survey, leads to different results. For example, \citealp{DiazGimenez+18} performed a comparison between CGs selected in the Main Galaxy Sample of the Sloan Digital Sky Survey Data Release 12 (SDSS-DR12) with a Hickson-modified criterion (HMCG)
and CGs identified by \cite{sohn+16} using a friends-of-friends algorithm (SHONCG). They found that only 44\% of the SHONCGs are also in HMCG sample. Recently, \cite{Zheng+20} identified a large sample of CGs with redshifts information also in SDSS-DR12 survey, using a relaxed Hickson-like criteria. The percentage of groups in common with previous samples is very poor and suggests that the population of CGs identified by different authors is different.

Although the usual criteria consider the local isolation of the groups, several studies indicate that some of the compact groups identified are not isolated but are associated with larger structures. \cite{mendel+11} found that 50\% of their sample of CGs were associated rich clusters\footnote{The nearest compact group is in the Virgo cluster \citep{Mamon89}.}, while \cite{DiazGimenez+15} found that 27\% of their CGs are embedded systems in redshift space.  
\cite{Zheng+21} noted that, in comparison with embedded CGs, isolated CGs have lower dynamical masses at given luminosity, which according to an evolution model of \cite{Mamon93} leads to more evolved systems.

Several studies using 3D real space information were performed to understand the nature of compact groups identified by different algorithms. Using mock galaxy catalogues, \cite{McConnachie+08} found that $30 \%$ of CGs selected in projection on the sky are physically dense in 3D space, while \cite{DiazGimenez&Mamon10} found that $\sim 65\%$ of CGs identified in redshift space are physically dense in 3D space, being the remaining chance alignments within groups or the field. These works studied the 3D nature starting from observational CGs. From another point of view, \cite{wiens+19}, using a particular set of semi-analytical galaxies, started from groups identified in 3D real space, and developed a criterion in 3D based on Hickson criteria. Their study focus in the evolution of the population of galaxies in CGs over cosmological time.

Recently, we have started a series of studies to analyse Hickson-like CGs using different semi-analytical tools. 
These studies make use of several publicly available semi-analytical model (SAMs) of galaxy formation and allowed us to perform two studies.
From the analysis of the frequency and nature of CGs identified in nine mock light-cones constructed from five different SAMs  (\citealt{DiazGimenez+20}, hereafter \citetalias{DiazGimenez+20}), we determine that the frequency and nature of CGs depend strongly on the cosmological parameters of the parent cosmological simulation. 
In Planck cosmologies \citep{PlanckCollaboration+20_cosmopars}, only roughly 35 per cent of the Hickson-like CGs are physically dense (as originally estimated by \citealt{Mamon86}), but using $N$-body cosmological simulations with higher mass and spatial resolution, this percentage is roughly half of all CGs.
We also found that intrinsic differences in the SAM recipes also lead to differences in the frequency and nature of CGs. 
On the other hand, analysing the formation history of Hickson-like CGs of four galaxies in light-cones built from different SAMs (\citealt{DiazGimenez+21}, hereafter  \citetalias{DiazGimenez+21}),  we identify four main channels of CG formation. 
Most CGs show late assembly, while only 10-20 per cent form by the gradual contraction of their orbits, and only a few per cent are early formed. We also observed a dependence of assembly class with the cosmology of the parent simulation: CGs assemble later in high-density universes. 
The comparison of observational properties of mock CGs indicates that early-assembling CGs are more evolved in terms of group compactness and galaxy luminosities.

The goal of this work is twofold.
Firstly, we propose a method to identify compact isolated dense groups in 3D real space (CG-3Ds) 
that reproduce the compactness of the original visual Hickson CG sample. 
Our procedure also includes the other two Hickson criteria: population and isolation. 
Secondly, we use this sample of 3D-CGs to analyse the ability of a Hickson-like algorithm \citep{DiazGimenez+18} to recover these systems in a flux limited catalogue. From this comparison, we intend to quantify the purity and completeness in the Hickson-like CG catalogues.
 
The layout of this work is as follows. 
In Section 2 we present the SAMs used in this work. 
In Section 3 we describe the criteria to identify CGs in 3D real space as well as the procedure to obtain a sample of those CGs that can be observed in a flux limited catalogue. 
In Section 4 we quantify the completeness and purity of the automatic Hickson-like catalogues by linking the sample of observable 3D-CGs and the sample of CGs identified using the automatic Hickson-like criteria. 
In Section 5 we evaluate the nature of the CGs  invented or missed by the Hickson-like algorithm. 
Finally, in Section 6 we summarise our results and present our conclusions. 

\section{Semi-analytical models of galaxy formation}
\label{sec:data}

\begin{table}
\setlength{\tabcolsep}{2pt}
\small{
\begin{center} 
\caption{Semi-analytical models analysed in this work \label{tab:sams}}
\begin{tabular}{clcccccc}
\hline
\hline
\multicolumn{1}{c}{SAM} &\multicolumn{4}{c}{Cosmology} & & \multicolumn{2}{c}{Simulation} \\ 
\cline{2-5} \cline{7-8}
acronym & \multicolumn{1}{c}{name} & $\Omega_{\rm m}$ & $h$ & $\sigma_8$ &  & box size & \# \\
\multicolumn{1}{c}{(1)\ \ } & \multicolumn{1}{c}{(2)} &\multicolumn{1}{c}{(3)} & (4) & (5) & & (6) & (7)  \\
\hline
G11 & \ WMAP1 & 0.25 & 0.73 & 0.90 &  & 500  & 6\,038\,533\\
G13 & \ WMAP7 & 0.27 & 0.70 & 0.81 &  & 500 & 5\,736\,985\\ 
H15 & \ Planck & 0.31 & 0.67 & 0.83 & & 480 & 4\,211\,108\\
H20 & \ Planck & 0.31 & 0.67 & 0.83 & & 480 & 5\,160\,118\\
A21 & \ Planck & 0.31 & 0.67 & 0.83 & & 480 & 3\,745\,859\\
\hline
\end{tabular}  
\end{center} 
\parbox{\hsize}{\noindent Notes:
The columns are:
(1): SAM: G11 
\citep{Guo+11}, G13 \citep{Guo+13}, H15 \citep{Henriques+15}, H20 \citep{Henriques+20} and A21 \citep{Ayromlou21}
(2): cosmology of parent simulation;
(3): density parameter of parent simulation;
(4): dimensionless $z$=0 Hubble constant of parent simulation;
(5): standard deviation of the (linearly extrapolated to $z$=0) power spectrum on the scale of $8\,h^{-1}\,\rm Mpc$;
(6): periodic box size of parent simulation [$\, h^{-1} \, \rm Mpc$];
(7): number of galaxies with stellar masses greater than $7\times 10^8 \, h^{-1} {\cal M}_\odot$. 
More accurate values for the simulations can be found at
 \url{http://gavo.mpa-garching.mpg.de/Millennium/Help/simulation}.}
\parbox{\hsize}{\noindent  $^{*}$ H15, H20 and A21 have smaller box size since they are run on a re-scaled version of the original Millennium simulation.}}
\end{table}

We analysed five SAMs run on the Millennium simulation (MS~I, \citealp{Springel+05}), which are listed below.
\vspace{-0.5\baselineskip}
\begin{itemize}
    \item \cite{Guo+11} (hereafter G11) run on the original MS~I with its WMAP1 cosmology  \citep{Spergel+03}. This model modifies earlier treatments of supernova feedback and galaxy mergers to agree better with observations of dwarf and satellite galaxies. The tracking of the angular momentum of different galaxy components allows that the size evolution of discs and bulges can be followed. Environmental processes such as tidal, ram-pressure stripping and merging are also applied.
    \item \cite{Guo+13} (G13) run on the MS~I with a WMAP7 cosmology \citep{Komatsu+11}. This model is almost identical to that run by \cite{Guo+11} but applied to a simulation run with different cosmological parameters.
    This re-scaling of the cosmological parameters of the MS~I is performed following the procedure of \cite{Angulo&White10}.
    \item \citeauthor{Henriques+15} (\citeyear{Henriques+15}, L-Galaxies, H15), run on the MS~I re-scaled to the Planck cosmology \citep{Planck+16}. 
    This model, which was fit to the observed stellar mass functions from $z=0$ to $z=3$, improves the representation of the build-up of the galaxy population over time, the present star formation activity of the low-mass galaxy population, and the environmental processes. For the latter, the authors suppressed ram-pressure stripping to solve the overestimation of the fraction of quenched satellite galaxies observed in \cite{Guo+11}.
    \item \citeauthor{Henriques+20} (\citeyear{Henriques+20}, L-Galaxies, H20), also run on the MS~I re-scaled to the Planck cosmology. This model maintains most of the improvements made by \cite{Henriques+15}, but the authors introduced modifications to spatially resolve phenomena happening in galactic discs by defining concentric rings where cold gas is partitioned into ${\rm HI}$ and ${\rm H_2}$ and converted into stars.
    \item \citeauthor{Ayromlou21} (\citeyear{Ayromlou21}, L-Galaxies, A21), again run on the MS~I re-scaled to the Planck cosmology. Based on the previous run of \cite{Henriques+20}, this new model introduces a novel gas stripping method. This model removes the ram-pressure stripping threshold of previous runs and applies this process to all galaxies. The model has also been improved by estimating the local environment of each galaxy using the simulation particles. Moreover, tidal stripping is applied to all satellite galaxies, regardless of their location. 
\end{itemize}
Table~\ref{tab:sams} quotes the different SAMs with their cosmological parameters and parent simulations.
Following \cite{Guo+11}, \cite{knebe15}, and \cite{irodotou19}, we only considered galaxies with stellar masses larger than $\sim 10^9 \, {\cal M}_\odot$ ($7\times 10^8 \, h^{-1} {\cal M}_\odot$) for the MS~I SAMs
(see queries to retrieve these data in the appendix of \citetalias{DiazGimenez+20}).

\section{Ideal compact groups in 3D real and redshift space}
\label{sec:criteria}

Historically, three key characteristics are associated with CGs: they are groups with a few members, in close proximity with one another, and they are relatively isolated in space. Under these conditions, they rise as crucial laboratories to study the effects of interactions between galaxies in over-dense environments which  have evolved relatively independently of the local and large scale environment. 

It is worth mentioning that most of the criteria used in the literature are applied on flux limited samples in redshift-space, with a few exceptions (see for instance, \citealp{wiens+19}).

\subsection{3D criteria}
\label{sec:criteria3D}
In this section, we describe the criteria adopted to identify groups in 3D real space that resemble as close as possible the idealistic view of CGs. Besides, with the aim of preserving the original idea of \cite{Hickson82}, we kept the main features of the visual \citeauthor{Hickson82}'s sample:
compactness, population and isolation.

\subsubsection{Compactness}
\label{sec:FoF-DG}

Since CGs are expected to be highly-dense systems, we define the compactness criterion based on the 3D separation between galaxies. 
We identify groups of galaxies by using a Friends-of-Friends (FoF) algorithm in 3D real space \citep{davis85}.

The FoF algorithm links galaxies that share common neighbours, i.e. it builds lists of galaxies (FoF halos) where each galaxy of the list fulfils that the physical separation with its nearest neighbour is below a given threshold. This threshold is known as the linking length, $r(z)$, which is defined as a dimensionless constant, $b(z)$, times the mean comoving intergalactic separation of the Universe, $1\,/\,\overline n^{1/3}(z)$ i.e.  
\begin{equation}
    r(z) = {b(z)\over\overline n^{1/3}(z)} = b(z)\,{L_{\rm box}\over N_{\rm gal}^{1/3}(z)} \ ,
    \label{eq:r0}
\end{equation} 
where $\overline n(z) = N_{\rm gal}(z)/L_{\rm box}^3$ is the mean comoving galaxy number density of the simulation at redshift $z$, and $L_{\rm box}$ is the redshift-independent comoving simulation box size.
\cite{Courtin+11} found the following empirical relation between the dimensionless linking length and the enclosed overdensity of virialised halos:  
\begin{equation} 
\label{eq:rho}
 b^{-3}(z) = b^{-3}_{f} 
\left(0.24 \frac{\Delta_{\rm vir}(z)}{178} + 0.68  \right) \ ,
\end{equation}
where $b_f$ is a fixed value\footnote{$b_f=0.2$ is usually adopted to identify halos of dark matter particles}, and $\Delta_{\rm vir} (z)$ is computed using eq.(16) of \cite{weinberg+03} assuming a $\Lambda$CDM cosmology. In this work, we only use the output of the simulation at $z=0$ and the cosmological parameters for each simulation (see Table~\ref{tab:sams}).

To ensure compactness similar to the original Hickson CGs, 
we must define a 3D linking length in such a way that, when looking at the groups in redshift space, the median projected separation between galaxy members is of the order of their sizes,
as it is the case in the observational catalogue of Hickson. In Appendix~\ref{app:r0}, we show how we determined a value of $r_0=r(z=0)= 90 \, h^{-1}{\rm kpc}$ to identify highly-dense groups in the periodic box of the simulations at $z=0$. 

\subsubsection{Population}
Although Hickson's original idea of population was to have groups with four or more members, when redshift measurements were incorporated to his sample, it was found that many of the groups were actually dense triplets, and much of the literature agreed to include those as compact groups. Therefore, among the highly-dense groups, we select those with three or more members (hereafter, FoF-DG).

Furthermore, the visual inspection performed by \cite{Hickson82} produced groups with galaxies similar in luminosity and sizes. 
Since FoF-DG groups are characterised by their very small inter-galaxy separations, and given that there is no restriction about the galaxy member stellar masses during the identification process, it is possible that a FoF-DG group is formed by very low-mass galaxies or even that there is a dominant galaxy with reasonable mass but its companions are very low-mass satellites. None of these two situations should be considered among the ideal systems that resemble the expected characteristics of Hickson-like compact groups. We introduce the following restrictions (hereafter ``mass population criterion''),

\begin{itemize}
\item The stellar mass of the most massive galaxy member has to be greater than $10^{10} \ h^{-1} \ {\cal M}_\odot$. As a reference, this limit implies that the dominant galaxy needs to be at least $40$ per~cent larger than the visible mass of the Triangle Galaxy (M33) in the Local Group (using the approximation of \citealt{corbelli03} and $h\sim0.7$).
This restriction excludes those FoF-DG groups 
of insufficiently massive galaxies, and sets a minimum expected stellar mass for the dominant galaxy. 
\item The stellar mass of the second most massive galaxy has to be greater than $10^{9.5} \ h^{-1} \ {\cal M}_\odot$. This limit was chosen to be approximately $50$ per~cent larger than the Large Magellanic Cloud, the largest galaxy satellite of the Milky Way (using the estimation of \citealt{marel06} and $h\sim 0.7$).
This requirement excludes FoF-DG groups that are just a host galaxy surrounded by their low mass satellites. 
\end{itemize}

\subsubsection{Isolation}
The other main characteristic of CGs is their isolation.
We are interested in highly-dense groups whose galaxies have little interaction with the environment. This condition ensures that CGs are physical entities by themselves, and everything that happens within them is due to the interactions between their galaxy members in a very confined portion of space.

To apply an isolation criterion, we follow two consecutive stages: first, we remove the highly-dense groups that are embedded in a larger structure (`loose group'); 
and then, we select from the remaining groups, those that can be considered as locally isolated (in 3D).

\begin{figure}
\centering
\includegraphics[width=0.4\textwidth]{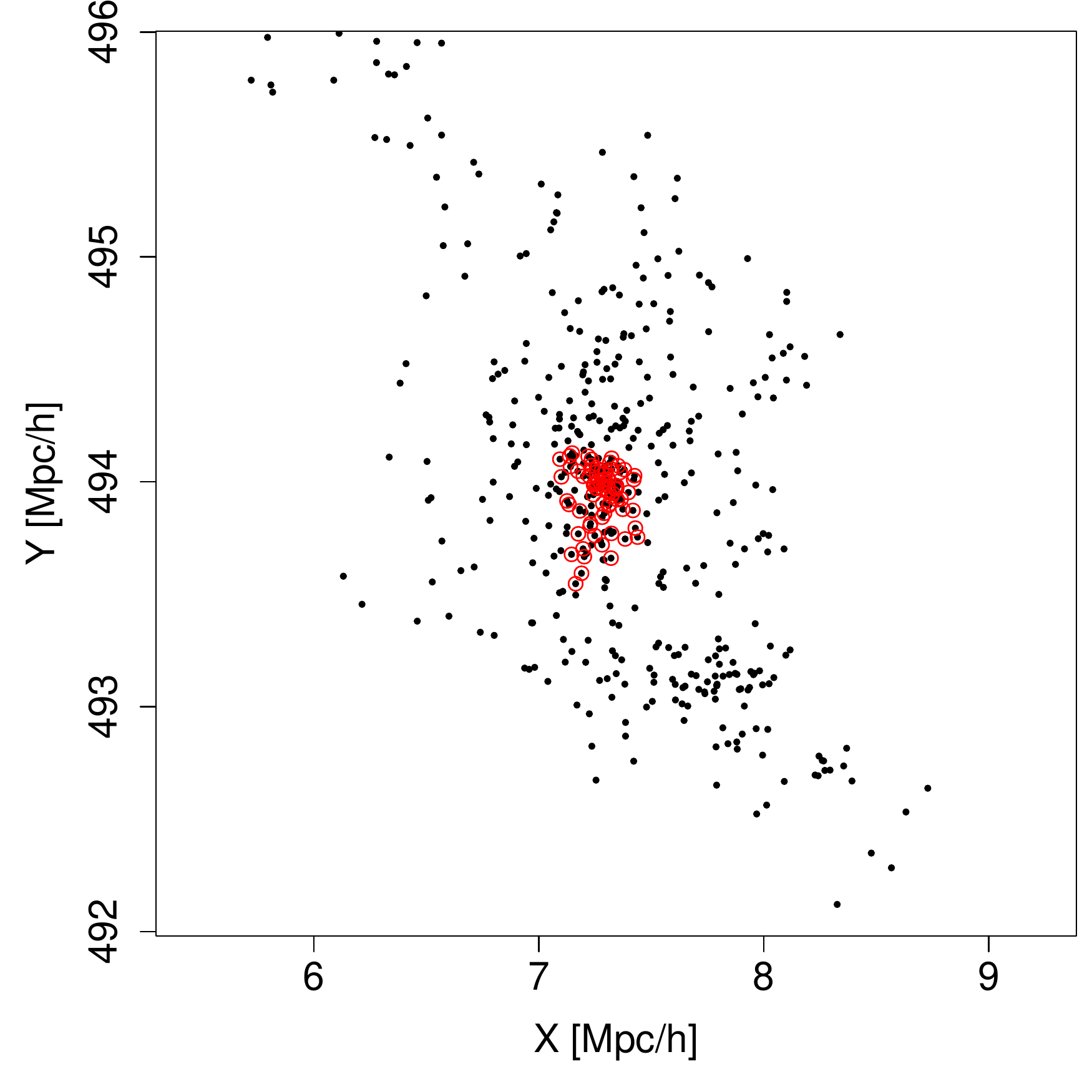}
\includegraphics[width=0.4\textwidth]{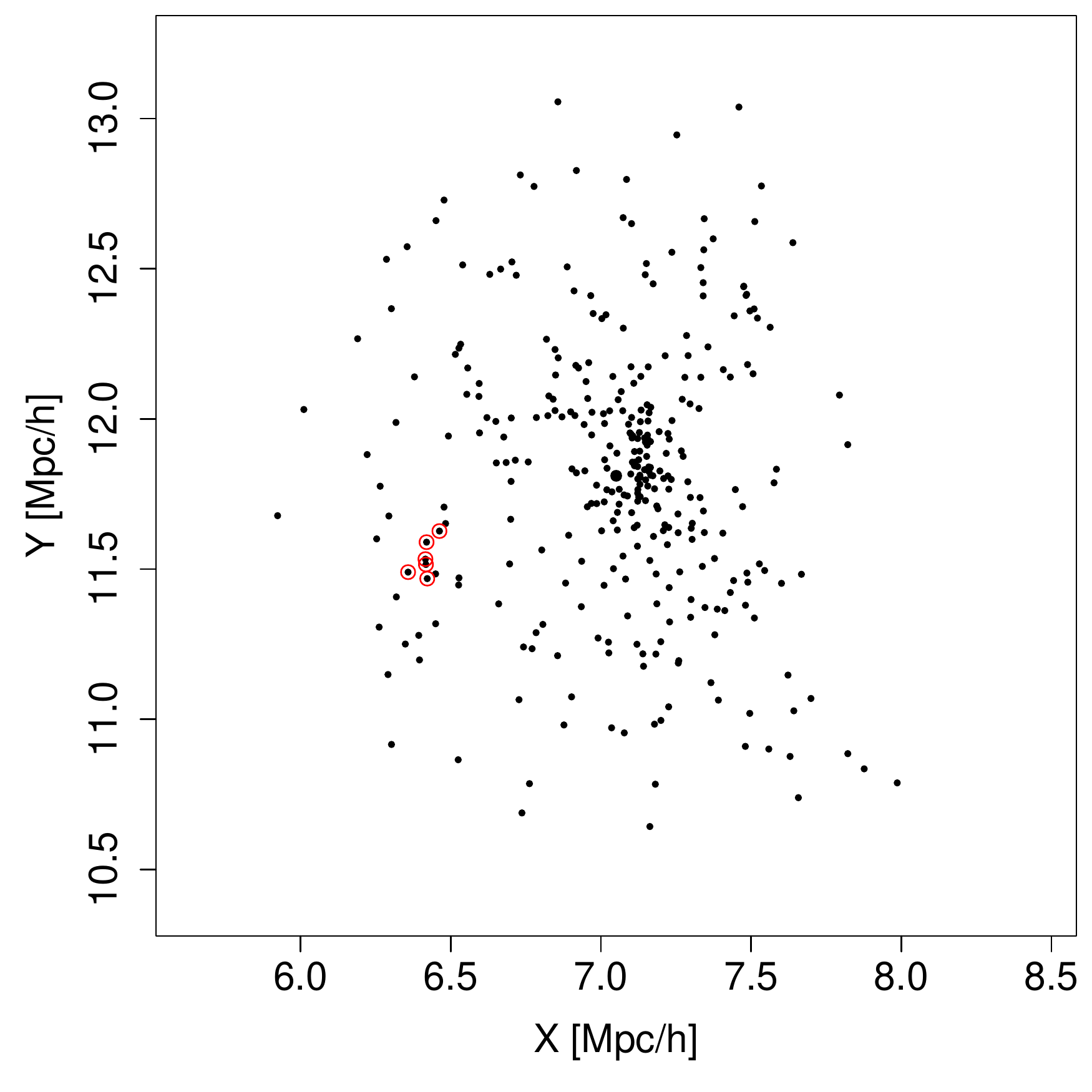}
\caption{Examples of FoF-DG groups that are embedded in Loose Groups extracted from the G11 SAM. These panels show the $x-y$ projections of galaxies in loose groups 
(\emph{black dots}), and galaxies in FoF-DGs (\emph{red open circles}).}
\label{fig:embed}
\end{figure}

\begin{description}
\item{\bf Global isolation: Non-embedded dense systems} 
\end{description}
We select highly-dense groups that are not embedded in other loose galaxy systems. 
To perform this selection, it is necessary to have a sample of virialized loose groups (LGs). 
We identify these systems using a FoF algorithm in real space at $z=0$ using a longer linking length $r_0$ appropriate to obtain virialized loose galaxy groups. 
Following previous studies \citep{Eke+04,Berlind+06,Zandivarez+14}, we adopt  $b_f=0.14$, and using the relations quoted in the eqs.~\ref{eq:r0} and~\ref{eq:rho}, we compute the values of $r_0$ to identify the samples of FoF-LGs in each SAM. The values of $r_0$ and number of FoF-LGs in each SAM are quoted in  Table~\ref{tab:groups_box}.

Then, we cross-match the samples obtained from both identifications, the FoF-LGs and the FoF-DGs, to obtain which FoF-DG group is embedded in an FoF-LG. It is worth noting that all members of a FoF-DG group will be members of a FoF-LG since the latter has been identified with a larger linking length (lower density). Therefore, by construction, a FoF-DG group is always embedded in a FoF-LG, but we are interested in finding whether the FoF-DG group is a substructure of a FoF-LG or if both can be considered as the same isolated galaxy system. 

To achieve this goal, we perform a member-to-member comparison and compute the ratio ${\cal R}$ between the number of members of the FoF-DG, $N_{\rm m}^{\rm FoF-DG}$, and the number of members of the FoF-LG, $N_{\rm m}^{\rm FoF-LG}$. 
The ratio ${\cal R}$ quantifies the relative size of the FoF-DG group in comparison with the FoF-LG:
${\cal R} \sim 0 $ indicates that the FoF-DG group is embedded in a much richer FoF-LG, 
while ${\cal R} \sim 1$ indicates that both groups can be considered as the same system. 
We consider a FoF-DG group as embedded in a FoF-LG when
\begin{equation*}
{\cal R} = \frac{N_{\rm m}^{\rm FoF-DG}}{N_{\rm m}^{\rm FoF-LG}} < 2/3
\end{equation*}
i.e., when the FoF-LG contains less than half as many extra members,
then both can be regarded as the same system.
Figure~\ref{fig:embed} shows two examples of FoF-DGs that have been discarded because they are embedded in FoF-LGs. In the top panel, the FoF-DG comprises the core of the FoF-LG, while in the bottom panel the FoF-DG is a substructure in the FoF-LG.

Nevertheless, for those FoF-DG groups that are considered identical to its corresponding FoF-LG and therefore not thought as embedded systems, we have also added another constraint that attempts to discard FoF-DG groups that are in the close vicinity of considerably large FoF-LGs. 
We calculate the 3D comoving distance of the FoF-DG system to the second closest FoF-LG, and then we compare it with the virial radius of the latter. Finally, we also classify a FoF-DG as embedded when the distance to the second closest FoF-LG is less than the virial radius of the FoF-LG, and the number of the galaxy members of the second closest FoF-LG is at least the double of the membership of the FoF-DG group, i.e., $Nm_{\rm 2nd-LG} > 2\, Nm_{\rm FoF-DG}$. \\

\begin{description}
\item{\bf Local isolation}
\end{description}
One of the attributes of the Hickson's CGs is that they are supposed to be locally isolated from other bright galaxies. 
Hickson's isolation criterion specifies that every group should satisfy that there is no other relatively bright galaxies (within a three magnitude range from the brightest member or brighter) inside a projected ring that extends from the angular projected group size to three times that distance. This approach was initially thought to avoid contamination of farthest galaxies in projection when redshift information was not available. But also, it eliminates the possibility of recent interactions with external bright galaxies in the evolutionary history of the group.

We therefore try to preserve this feature by introducing a criterion to produce a sample of locally isolated FoF-DG groups, this time in 3D space.
We then consider a FoF-DG group as locally isolated when the there is no other galaxies with stellar masses greater than $10^{9.5} \ h^{-1} \ {\cal M}_\odot$ inside a spherical shell of internal radius $r_{\rm max}^{\rm FoF-DG}$ equal to the maximum 3D comoving distance of the group members to the geometrical centre, and external radius three times that distance\footnote{The mass threshold is the minimum stellar mass allowed for the second most massive galaxy member imposed in the population criterion.}. 

We apply this criterion only to those groups that survived the global isolation filter.

\begin{table}
\setlength{\tabcolsep}{3pt}
\small
\centering
\caption{Galaxy groups identified for each SAM. 
\label{tab:groups_box}}
\begin{tabular}{l|rr|rr|rr|rr|rr}
\hline
& \multicolumn{10}{c|}{SAMs} \\
\cline{2-11}
& \multicolumn{2}{c|}{G11} &  \multicolumn{2}{c|}{G13} & \multicolumn{2}{c|}{H15} & \multicolumn{2}{c}{H20} &  \multicolumn{2}{c|}{A21} \\
\hline
\multicolumn{11}{c}{\bf Simulation Box}\\
\hline
\underline{3D LG} & &&&&&&& \\ 
$r_0$ [$h^{-1} \ \rm kpc$] & \multicolumn{2}{r|}{363}  & \multicolumn{2}{r|}{372} & \multicolumn{2}{r|}{401} & \multicolumn{2}{r|}{375} & \multicolumn{2}{r|}{417} \\
FoF-LGs & \multicolumn{2}{r|}{298\,985}  &  \multicolumn{2}{r|}{304\,265} & \multicolumn{2}{r|}{219\,474} & \multicolumn{2}{r|}{260\,760}& \multicolumn{2}{r|}{138\,013} \\
\hline
\underline{3D DG} & &&&&&&& \\ 
$r_0$ [$h^{-1} \ \rm kpc$] & \multicolumn{2}{r|}{90}   & \multicolumn{2}{r|}{90} & \multicolumn{2}{r|}{90} & \multicolumn{2}{r|}{90} & \multicolumn{2}{r|}{90}  \\
FoF-DGs   & \multicolumn{2}{r|}{171\,500}  & \multicolumn{2}{r|}{150\,173} &  \multicolumn{2}{r|}{77\,458} & \multicolumn{2}{r|}{129\,339} & \multicolumn{2}{r|}{24\,131} \\ 
\hline
3D-CGs & \multicolumn{2}{r}{41\,053} &  \multicolumn{2}{r}{35\,304} & \multicolumn{2}{r}{19\,695} & \multicolumn{2}{r}{25\,551}  & \multicolumn{2}{r}{3\,901} \\
\hline
\multicolumn{11}{c}{\bf Mock catalogue} \\
\hline
z-CG$^{+1}$ & \multicolumn{2}{r|}{94\,679}  & \multicolumn{2}{r|}{7\,788} & \multicolumn{2}{r|}{49\,999} & \multicolumn{2}{r|}{74\,669} & \multicolumn{2}{r|}{12\,139} \\
z-CG$^{+3}$ & \multicolumn{2}{r|}{4\,134 }  & \multicolumn{2}{r|}{3\,527}  & \multicolumn{2}{r|}{2\,615}  &  \multicolumn{2}{r|}{3\,827} & \multicolumn{2}{r|}{702}  \\ 
z-CG$^{+4}$ & \multicolumn{2}{r|}{927}  & \multicolumn{2}{r|}{625}  & \multicolumn{2}{r|}{389}  &  \multicolumn{2}{r|}{797} & \multicolumn{2}{r|}{66}  \\
\hline
\hline
HMCG$^{+3}$  & \multicolumn{2}{r|}{10\,734}  &   \multicolumn{2}{r|}{6\,403} &   \multicolumn{2}{r|}{5\,876} & \multicolumn{2}{r|}{7\,711}  & \multicolumn{2}{r|}{3\,351} \\
HMCG$^{+4}$ & \multicolumn{2}{r|}{2248}  &   \multicolumn{2}{r|}{1213} &   \multicolumn{2}{r|}{972} & \multicolumn{2}{r|}{1446}  & \multicolumn{2}{r|}{546} \\
\hline
\end{tabular}
\parbox{\hsize}{\noindent Notes: 
FoF-LG: loose groups identified in 3D-space with linking length quoted as $r_0$; 
FoF-DG: over-dense groups identified in 3D-space with linking length quoted as $r_0$; 
3D-CGs: ideal compact groups obtained after applying the mass population and isolation criteria to the FoF-DG.  
z-CG: ideal compact groups observed in a flux limited all-sky mock catalogue (redshift space).  
HMCG: groups identified in a flux limited all-sky mock catalogue with an automatic Hickson-like algorithm. 
Superscripts indicate the lower limit of group members in each sample.\\
}
\normalsize
\end{table}

\subsection{The identification of 3D-CGs}
\label{3dcg}

In each SAM, we identify CGs in 3D real space by applying the compactness, population and isolation criteria described above.  

In Table~\ref{tab:groups_box} (Simulation Box section), 
for each SAM we quote the number of groups identified.
It is worth bearing in mind that each criterion is applied only to those groups that passed through the restrictions of the previous criterion (i.e., in the following order: compactness, population, and isolation). 
We quote the number of FoF-DG groups having three or more members identified using a linking length $r_0=90 \  h^{-1} {\rm kpc}$ (compactness criterion). After applying the population and isolation criteria, we obtain our ideal sample of compact groups in 3D, labelled as 3D-CGs. 

We find that the fraction of groups discarded for not accomplishing the mass population criterion
is between 19 to 37 per~cent (22, 25, 19, 37 and 34 \% of the FoF-DG groups for G11, G13, H15, H20, and A21 respectively).
Among those that fulfilled the population criterion, at least 2/3 of 
groups are embedded systems ($> 66\%$) and, therefore, are discarded in our work.
Finally, when the local isolation criterion is applied on the surviving sample, 
roughly 8 per~cent of the remaining sample is discarded (10, 9, 6, 7, and 6\% for G11, G13, H15, H20, and A21, respectively).
Therefore, the global isolation criterion has the largest impact at restricting the sample of compact isolated dense groups.  

All in all, the final sample of 3D-CGs represents between $16 \%$ to $25 \%$ of the original FoF-DG groups (24, 24, 25, 20 and 16\% for G11, G13, H15, H20 and A21, respectiveyly), and this is the sample we consider as the ideal compact groups in 3D in each SAM. 

\subsubsection{Comparison with the \citeauthor{wiens+19} criteria}

Similar to our work,  \cite{wiens+19} devised a method to find compact groups in 3D real space to be applied to the semi-analytical galaxies of the \cite{DeLucia&Blaizot07} SAM. They also followed the main features of Hickson's criteria: compactness, population and isolation. 
We compare the results using our and those of \citeauthor{wiens+19}.

Briefly, the criteria of \citeauthor{wiens+19}  involve 
identifying overdense groups with three or more members using a DBS{\small CAN} algorithm \citep{dbscan} with a neighbourhood parameter $NH=50 \ h^{-1} \rm kpc$, 
removing systems that are not isolated in terms of the stellar mass density enclosed in a sphere of radius $1 \, h^{-1}\rm Mpc$ (SR criterion), 
eliminating groups with one single dominant galaxy (MR criterion), 
excluding dwarf galaxies with stellar masses lower than $5 \times 10^8 \, h^{-1}\ {\cal M}_\odot $,
and filtering galaxies in radial velocity located at more than $1000 \, \rm km\,s^{-1}$ from the median velocity of the group centre. 

To perform the comparison, we use the G11 SAM and start with the FoF-DG groups identified with a linking length of $r_0=90 \, h^{-1} \,\rm kpc$ (Sect. \ref{sec:FoF-DG}), Then, we investigate the differences between our selection of compact and isolated groups and those we could obtain whether we apply the SR and MR criteria of \citeauthor{wiens+19}. We will not consider the two last criteria imposed by those authors since the stellar mass resolution of the MS-I simulation already avoids including galaxies below the limit proposed by \citeauthor{wiens+19}, and the radial velocity filter is not needed when working in 3D real space.  

To start with, on the sample of $171\,500$ FoF-DG groups we apply our mass population criterion onto the two most massive galaxies of the groups (${\cal M}_1 > 10^{10}\, h^{-1}\,{\cal M}_\odot$ and ${\cal M}_2 > 10^{9.5}\, h^{-1}\,{\cal M}_\odot$). On the other hand, we apply the \citeauthor{wiens+19} MR criterion which involves the three most massive galaxies in the groups ( $({\cal M}_2 + {\cal M}_3)/{\cal M}_1 > 0.1$). Our criterion leads us to $134\,030$ groups (78\%), while the \citeauthor{wiens+19} criterion produces $146\,439$ groups (85\%). To investigate the differences between these two samples, we analyse the fraction of groups that satisfy both criteria or only one of them.
\begin{itemize}
    \item $73\%$ of the FoF-DG groups
    fulfil both criteria at the same time.
    \item When applying the \citeauthor{wiens+19} MR criterion to the sample of $134\,030$ groups that survived our low stellar mass criterion, we find that  $6\%$ 
    of the groups do not satisfy their criterion.
    \item When applying our mass population criterion to the sample of $146\,439$ groups that survived the \citeauthor{wiens+19} MR criterion, we find that $15\%$ 
    of the groups do not satisfy our criterion.
\end{itemize}

Finally, we analyse the final samples of compact and isolated 3D compact groups produced by the two methods.
Besides the criteria on stellar masses, both methods require isolation. 
On our side, our selection involves avoiding embedded systems and also asking for local isolation, as explained above. On the other side, \citeauthor{wiens+19} include an isolation criterion in terms of the logarithm of the ratio between the stellar mass density within a shell of external radius $1 \, h^{-1}\, \rm Mpc$ and internal radius defined by the maximum distance of a group member to the group centre, and the stellar mass density within the sphere defined by the distance to the farthest galaxy member. Their chosen default threshold for this ratio is $-4$. 

All our criteria applied to the sample of FoF-DG lead us to $41\,053$ CG-3Ds, while the \citeauthor{wiens+19} criteria produce $68\,478$ final groups. On these resulting samples we find that: 
when applying the \citeauthor{wiens+19} criteria to our sample of CG-3Ds, $79\%$ of the groups also fulfil the \citeauthor{wiens+19} criteria. On the other hand, when applying our criteria to the $68\,478$ that satisfied all \citeauthor{wiens+19} criteria,  $47\%$ of the groups are selected. 

The criteria presented in this work are more restrictive than the \citealp{wiens+19} definition. The more important difference is produced by the isolation criteria. A large fraction of groups considered as isolated by \citeauthor{wiens+19} do not survive our selection.

\subsection{Observing 3D-CGs in Redshift Space}
\label{sec:3Dmock}

Once we selected our galaxy sample of ideal compact isolated dense groups, we would like to compare it to a sample of CGs identified in redshift space using the automatic Hickson-like criteria. 

To find a correlation between both kinds of groups, we first need to transform the 3D-CG sample into a sample of systems that can be observed in a flux limited redshift space catalogue. 

We therefore place an observer and transform the 3D real space coordinates and properties, into observable parameters. We basically follow the usual steps for building a mock catalogue from a simulation box, but using only the galaxies in groups:

\begin{itemize}
    \item We choose to mimic the observational features of the SDSS Main Galaxy Sample, i.e., a flux limited redshift survey with an apparent magnitude limit in the $r$-band observer frame of $17.77$.
    \item Each periodic simulation box is replicated in each 3D direction to reach a distance of $\sim 1 \, h^{-1} \rm \, Gpc$ (this implies a maximum redshift of $\sim 0.35$, similar to that observed for the most distant objects in the SDSS Main Galaxy Sample). 
    In this manner, the observer is placed at the centre of a super-box of $\sim 2 \, h^{-1} \rm \, Gpc$ of side (4 boxes replications in each direction) in order to build an all-sky catalogue. For the purposes of this work, we only consider the $z$=0 snapshot of the simulation when building the mock catalogues.  
    \item Using the 3D comoving  positions, peculiar  velocities,  and  rest-frame  absolute magnitudes, we compute the angular positions, redshifts, and apparent magnitudes\footnote{k-corrections were inferred from the de-kcorrection procedure described in \cite{DiazGimenez+18}} for each galaxy member of the 3D-CGs.
    \item We select galaxies within a sphere of comoving radius of  $ 1 \, h^{-1} \rm \, Gpc$.
    \item Finally, we keep galaxies with $r$-band observer frame apparent magnitude less than $17.77$. It should be noted that once this magnitude cut is introduced, our samples are almost completely confined to 1 simulation box from the observer.
\end{itemize}

In Table~\ref{tab:groups_box} (Mock catalogue section), we quote the number of 3D-CGs in redshift space that comprise at least one, three or four members brighter than the apparent magnitude limit (hereafter, z-CG$^{+1}$, z-CG$^{+3}$ and z-CG$^{+4}$, respectively). 
In future sections we will be particularly interested in those 3D-CGs that are plausible of being identified with standard algorithms in observations, i.e. those that have 3 or more members brighter than the apparent magnitude limit.

Less than $10$ per~cent of the 3D-CG that are included in the sphere of $ 1 \, h^{-1} \rm \, Gpc$ have at least one galaxy observable in a flux limited all-sky catalogue. 
Taking into account that the search of CGs in observational catalogues has a minimum number of members of $3$ or $4$ galaxies, the percentage of 3D-CGs that remain ``observable'' in those conditions is very small ($\lesssim 0.5\%$).

The small chance of observing a 3D-CG in a flux limited catalogue is given by an observational restriction that is not possible to overcome. In accordance with our results, for a catalogue such as the SDSS Main Galaxy Sample, with a solid angle of $0.91\pi$ \citep{DR12b}, one should not expect to find more than between $\sim 160$ and $\sim 940$ true CGs with three or more members brighter than $r=17.77$, depending on the SAM.  
In our previous works, we have worked with CGs with four or more members. According to these results, the SDSS Main Galaxy Sample should contain between 15 and 211 CGs with at least four members. 


\section{Recovering 3D-CGs from observations}
In this section, we investigate the ability of the standard observational redshift-space algorithm to find the ideal CGs. 

\subsection{Hickson-like CG catalogue}
\label{sec:hcg}
We build an all-sky mock catalogue of galaxies in redshift space following the procedure described in Sect.~\ref{sec:3Dmock}, but this time using all the galaxies in one simulation box. 

We identify CGs in redshift space in a similar manner to that described in \citetalias{DiazGimenez+20}. Briefly, the algorithm searches for groups whose galaxies satisfy the population, compactness, isolation, flux limit of the brightest galaxy and velocity concordance criteria, always considering galaxies in a range of three magnitudes from the brightest. 

In this work, we introduced two modifications to the original algorithm presented in \cite{DiazGimenez+18} and used in \citetalias{DiazGimenez+20}. 
First, instead of identifying CGs with at least four members, we allow the algorithm to include triplets in our sample. Second, in those cases where a group of fewer galaxies inside a CG also satisfies all the CG criteria (CG inside a CG), we now save the smaller association of galaxies. This is achieved by running the algorithm starting from the lowest number of members instead of starting from the maximum allowed number of members as we did in our previous works. We will name these samples obtained using  Hickson-like criteria as HMCG$^{+3}$ (Hickson Modified Compact Groups, since they are obtained from the modified algorithm presented in \citealt{DiazGimenez+18}).

The number of HMCGs identified in each SAM is quoted in the last row of Table~\ref{tab:groups_box}. In \citetalias{DiazGimenez+20}, we have investigated the variations from SAM to SAM of the occurrences of CGs as a function of the simulation underlying cosmologies and the particular recipes of SAMs. We concluded that both have an impact on the number of CGs that can be found in these catalogues. In that work, the samples from H20 and A21 had not been analysed, so these new results are showing how the changes introduced from H15 to H20 and A21 are impacting on the occurrence of CGs.

\subsection{Purity and completeness of the Hickson-like catalogue }

Using our ideal observable sample z-CG defined in Sect.~\ref{sec:3Dmock}, we investigate the purity and completeness of the HMCG sample. These two properties of a catalogue are useful to test the finding algorithms. If the purity is high, the sample of groups is reliable and has little contamination. If the completeness is high, it means that the algorithm is able to find most of the true systems.
Let $N_{\rm R}$ be the number of recovered CGs, i.e. HMCGs that match z-CGs.
We  then define: 
\begin{description}
\item[\bf Purity:] the percentage of HMCGs that are also true  z-CG: $$P=100\frac{N_{\rm R}}{N_{\rm HMCG}}$$
\item[\bf Completeness:] the percentage of z-CG that are also found by the observational algorithm:
$$C=100\frac{N_{\rm R}}{N_{{\rm z\hbox{-}CG}}}$$  
\end{description}

To find the $N_{\rm R}$ groups that belong to both samples (observational and ideal), we perform a member-to-member matching process. In a first stage, if at least one galaxy member of an HMCG is also a member of a z-CG, then we considered such group as a match. The number of matching groups is quoted in Table~\ref{tab:p&c}.
Given that our definition of a matching group is quite permissive (at least one matching member), we analyse the degree of coincidences. We then quantify the number of matched HMCG that have all their galaxy members in a single z-CG. 
Regardless of the SAM, we find that roughly $83\%$ of the recovered HMCGs share all their members with the associated z-CG.

Using the number of matching groups, we analyse the purity and completeness of the sample HMCG$^{+3}$ with respect to z-CG$^{+3}$. The values of the purity and completeness of the HMCG$^{+3}$ sample are quoted in Table~\ref{tab:p&c}.
The purity of the HMCG$^{+3}$ catalogues is lower than $11 \%$ in all of the SAMs. This means that the samples of HMCG$^{+3}$ are highly contaminated by spurious identifications. 
On the other hand, the completeness of HMCG$^{+3}$ catalogues is also poor (lower than $24\%$). Therefore, the observational algorithm is missing most of the true CGs. 

In several previous studies, the minimum number of members in compact groups was four galaxies. Therefore, we also computed the purity and completeness of the observational sample HMCG$^{+4}$ relative to the z-CG$^{+4}$. We found that these percentages are even lower than those calculated before. Purity reaches at most 6\% (5\%, 6\%, 5\%, 6\%, and 2\% for G11, G13, H15, H20, and A21, respectively), while completeness drops to $\sim 12\%$ (13\%, 12\%, 13\%, 11\%, and 12\% for G11, G13, H15, H20, and A21, respectively). Even if we compute the purity and completeness of the HMCG$^{+4}$ relative to the z-CG$^{+3}$, the purity reaches at most $8\%$ and the completeness of observational sample drop to $4\%$.

Therefore, at best (using three or more galaxy members), $\sim 90\%$ of the HMCG are spurious identifications, and $\sim 80\%$ of the ideal z-CG are missed. In the following sections we will investigate the nature of the spurious systems and the reasons why the automatic algorithm cannot find the ideal groups. 

\begin{table}
\setlength{\tabcolsep}{3pt}
\centering
\caption{Purity and completeness of the HMCG$^{+3}$ relative to the z-CG$^{+3}$ 
\label{tab:p&c}}
\begin{tabular}{rrrrrr}
\hline
&  G11  & G13  & H15 & H20 & A21 \\
\hline
$N_{\rm R}$       & 979   & 691 & 578 & 839 & 152 \\
Purity        &  9\%   &  11\% &  10\% & 11\% & 5\% \\
Completeness  &  24\%  &  20\%  & 22\% & 22\% & 22\% \\
\hline
Spurious HMCG$^{+3}$ &  9755  & 5712 & 5298 & 6872 & 3199 \\
 Missing z-CG$^{+3}$ & 3156  & 2836 & 2037 & 2990 & 550 \\
\hline
\end{tabular}
\end{table}

\section{Spurious and missing CGs}

In this section, we focus on understanding the reasons why the purity and completeness of the HMCG$^{+3}$ catalogues are so low.

\subsection{Spurious HMCG$^{+3}$}
\label{sec:spurious}

Firstly, we study the HMCG$^{+3}$ without a counterpart in the z-CG$^{+3}$ sample (hereafter, spurious). These groups contaminate the HMCG$^{+3}$ sample causing the purity to drop. 
The number of spurious\footnote{The number of spurious HMCG$^{+3}$ is $N_{\rm HMCG^{+3}}-N_{\rm R}$.} HMCG$^{+3}$ in each SAM is quoted in Table~\ref{tab:p&c}.
Roughly $90$ per~cent of the HMCG$^{+3}$ are spurious in each SAM.

We wonder whether the spurious groups could be associated 
with FoF-DG groups that failed mass population or the isolation criteria in 3D to be 3D-CGs, or 
with 3D-CGs that failed to be observable with three or more members in the flux limited catalogue to be z-CG$^{+3}$, but still the Hickson-like algorithm finds them as HMCG$^{+3}$. 

Therefore, we examine whether the members of the spurious HMCG$^{+3}$ are also members of any FoF-DG groups (Sect.~\ref{sec:FoF-DG}). In this way, we associate the spurious HMCG$^{+3}$ with a highly overdense FoF-DG structure, when possible. 

We split the spurious HMCG$^{+3}$ groups into four categories depending on the number of matched members with a FoF-DG group:

\begin{description}
\item [\bf FM] (Full Match): All of the members of a spurious group match with members of a FoF-DG group.
\item [\bf MM$+$] (Medium Match $+$): More than 50\% of the members of the spurious group match with members of a FoF-DG  group.
\item [\bf MM$-$] (Medium Match $-$): Less than 50\% of the members of the spurious group match with members of a FoF-DG group.
\item [\bf NM] (No Match): No member matches. The spurious group does not have an associated FoF-DG group. 
\end{description}

\begin{figure}
\includegraphics[width=0.5\textwidth]{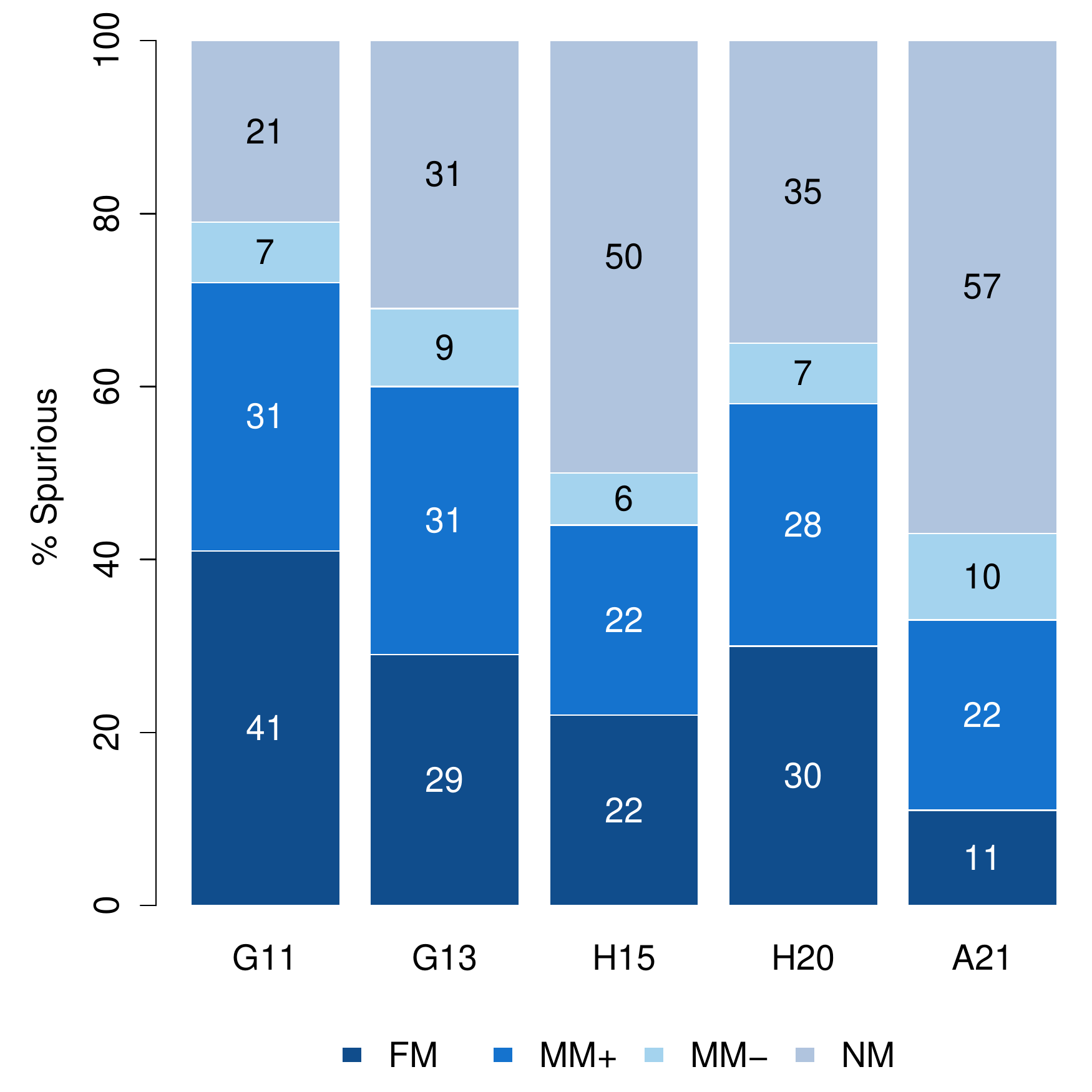}
\caption{Percentage of spurious HMCG$^{+3}$ within each category split according to the number of members that inhabit a FoF-DG group (see Sect.~\ref{sec:spurious}).
}
\label{fig:barplot_norec+3}
\end{figure}

In Fig.~\ref{fig:barplot_norec+3}, we show the percentages of spurious HMCG$^{+3}$ groups that belong to each category, for each SAM.

We find that between 21 to 57\% (depending on the SAM) of the spurious HMCG$^{+3}$ are not linked to any FoF-DG group (NM). Therefore, the observational algorithm is producing on average one third of HMCG$^{+3}$ that are not as dense as expected. 
We can go one step further and investigate whether these NM systems are at least inhabiting FoF-LGs. After performing this new cross-match procedure, we obtained that $\sim 89\%$ of the NM systems are embedded in a FoF-LG (88, 90, 88, 87, and 89 per cent in G11, G13, H15, H20, and A21, respectively). The remaining $\sim 11\%$ of NM systems can be considered as chance alignments in the field or filaments. 

Regarding the spurious HMCG$^{+3}$ that are indeed related to a FoF-DG, the FM and the MM$+$ categories comprise groups where most of their members inhabit in a unique FoF-DG structure, and this accounts for 72, 60, 44, 58, and 31\% of the HMCG$^{+3}$ in G11, G13, H15, H20, and A21, respectively. Therefore, it seems very likely that most of the observable members of these spurious HMCG$^{+3}$ are in the same FoF-DG, plus few interlopers.

To further deepen the analysis, we study what happens with the spurious groups related to a FoF-DG group (FM, MM+ and MM-), but its associated FoF-DG failed to be a 3D-CG or failed to be observed as a z-CG$^{+3}$. 
 
There are several stages where a FoF-DG could have been either rejected to be considered a 3D-CG (mass population or isolation) or not observed in a flux limited catalogue. The possibilities are as follows:
\begin{itemize}

\item the FoF-DG group is a large galaxy with its satellites or it is a system of only dwarf galaxies (mass population criterion)
\item the FoF-DG group is actually embedded in an FoF-LG (global isolation criterion)
\item the FoF-DG group is not locally isolated (local isolation criterion)
\item the 3D-CG has less than 3 members with apparent magnitude brighter than the catalogue limit (it may have 1 or 2 galaxies with $r<17.77$). 
\end{itemize}

In Table~\ref{tab:spurious} we show the percentage of spurious groups that are associated with a FoF-DG group ($N_{\rm FoF-DG}^{\rm sp}$), but the associated FoF-DG has failed to be a 3D-CG because it does not pass the mass population or the isolation criteria, or being a 3D-CG it failed to be observable with three or more members after the flux limit restriction. Each column shows the percentage rejected in each restriction.   
Regardless of the SAM, 
most of the associated FoF-DG groups are actually embedded within loose groups (near 88\%), and therefore they were rejected as 3D-CGs by the local isolation criterion. 
It means that the automatic Hickson-like algorithm (redshift-space) identifies a considerable fraction of (spurious) HMCG$^{+3}$ that are indeed highly-dense substructures but inhabiting larger loose groups. It might indicate that the observational isolation criterion is deficient.

\begin{figure}
\includegraphics[width=0.5\textwidth, trim= 120 260 200 100,clip]{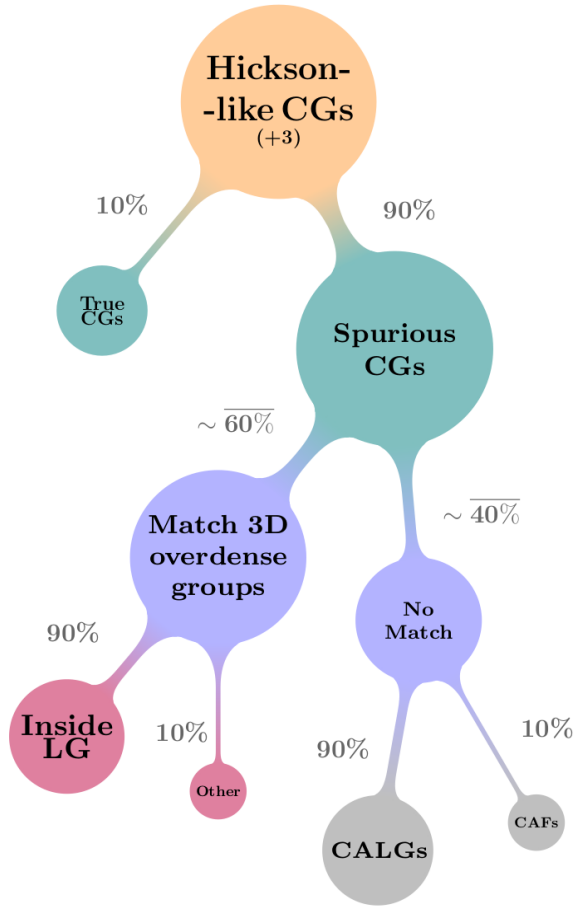}
\caption{Schematic drawing of the purity of the automatic Hickson-like CGs sample. The percentages of each branch are consistent regardless the SAM, except for the passage from 2nd to 3rd level (cyan to violet bubbles), where the quoted values are  obtained averaging over all the SAMs. Red bubbles at the third level indicate the percentages of 3D overdense groups associated to the spurious groups that did not achieve the global isolation criterion, or failed the other criteria (mass population, local isolation, or membership after the flux limit). Gray bubbles indicate whether CG members are embedded in loose groups (CALGs) or they are chance alignments in the field or filaments (CAFs).
}
\label{fig:bubbles}
\end{figure}

Therefore, from these results, we can conclude that, averaging for all the SAMs, the automatic Hickson-like criteria to identify CG in observational catalogues produce samples where only 10\% are actually very dense isolated groups, 55\% are very dense non-isolated systems, 32\% are systems not as dense as expected (chance alignments within loose groups), and 3\% are chance alignment in the field or filaments (see Fig.~\ref{fig:bubbles}).

Finally, in previous studies (e.g., \citetalias{DiazGimenez+20,DiazGimenez+21}), we used to split the samples of HMCG$^{+4}$ into physically dense \nobreak{(Reals)} and chance alignments (CAs). 
This classification was originally defined by \cite{DiazGimenez&Mamon10}, and it is based on 3D real space information of Hickson-like CGs identified in redshift space. 
To assess the reliability of the \cite{DiazGimenez&Mamon10} criterion, we applied it to our sample of HMCG$^{+3}$ and compared the results with the new information obtained for the spurious sample of HMCG$^{+3}$ in this section.  
We observe that the percentage of the formerly called Reals HMCG$^{+3}$ is a conservative estimate (lower limit) of the groups that are actually associated with truly dense structures in 3D real space. We also found that these estimates are not free from contamination, since $9-32\%$ (depending on the SAM) of the systems called Reals are not highly-dense groups. A full description of these results is shown in Appendix~\ref{app:reals}. 

\begin{table}
\setlength{\tabcolsep}{2pt}
\centering
\caption{Percentages of spurious groups associated with a FoF-DG group that has failed to be considered a 3D-CG or it failed to be observable with three or more members after the flux limit.
\label{tab:spurious}}
\begin{tabular}{cccccc}
\hline
SAM & $N_{\rm FoF-DG}^{\rm sp}$ & Mass & Global & Local & Flux \\
& & population & isolation  & isolation & limit \\
\hline
 G11 & 7679 & 3\%& 91\% & 4\% & 2\% \\
 G13 & 3908 & 6\% & 88\% & 4\% & 2\% \\
 H15 & 2636 & 6\% & 89\% & 2\% & 3\% \\
 H20 & 4447 & 12\% & 83\% & 3\% & 2\% \\
 A21 & 1386 & 8\% & 90\% & 1\% & 1\% \\
\hline
\end{tabular}
\end{table}

\subsection{Missing z-CG$^{+3}$}

We now study the $\sim$ 80~per~cent of ideal z-CG$^{+3}$ groups not found by the automatic Hickson-like algorithm in redshift space (missing groups). 

The numbers of missing groups\footnote{The number of missing z-CG$^{+3}$ is not exactly $N_{{\rm z-CG^{+3}}}-N_{\rm R}$ since there are a few HMCG$^{+3}$ inhabiting more than one z-CG$^{+3}$.} are quoted in Table~\ref{tab:p&c}.  
Between 76 and 80 per~cent (depending on the SAM) of the z-CG$^{+3}$ are not identified by the observational algorithm. 

To answer the question of why the algorithm does not recover all z-CG$^{+3}$, we study which of the Hickson-like criteria is not fulfilled by these groups. 
We find that most of the z-CG$^{+3}$ ($>99\%$, regardless of the SAM) do not fulfil 
either 
the restriction of the brightest galaxy ($m_{\rm bri}<m_{\rm lim}-3$), i.e., their brightest galaxy is fainter than 14.77, 
or they have less than 3 members in a range of 3 magnitudes from the brightest.
The very few remaining missing groups do not fulfil the other criteria, compactness, isolation or velocity concordance criteria. 
In conclusion, the main reason to miss many of the ``ideal'' CGs (z-CG$^{+3}$) is the observational restriction of considering members in a range of 3 magnitudes from the brightest (which has also forced us to include in the algorithm the flux limit to the brightest galaxy). 
Such restriction is important for computing population and isolation in the Hickson-like algorithm.

\section{Summary}
\label{sec:discus}

In this work, we studied the nature of CGs, starting from a different starting point than previous studies. We defined what should be considered as a CG in 3D real space and from there, we analyse how good the automatic Hickson-like criteria are at recovering such a sample of CGs, which we consider ideal.
Therefore, as a first step, we designed a set of criteria for identifying CGs in real space that was applied to five different SAM samples. 
To define what we consider as a 3D compact group, we took as a guide the global characteristics used by \cite{Hickson82}: compactness, population and isolation, but considering all properties in real space. 
We started with the identification of highly-dense groups using a FoF percolation algorithm \citep{davis85} with a considerably higher density contrast than usually used for normal galaxy groups. This high density contrast produces groups that, when observed in redshift space, present separations between galaxies similar to the median separation of galaxies in the original visual observational \citeauthor{Hickson82} sample. 
Once these physically dense groups were identified, we eliminated from the samples of dense groups those that are formed by a massive galaxy and its low-mass satellite galaxies. Then, we selected those that are not substructures of loose groups. 
Finally, we selected groups that are locally isolated from massive galaxies in their near surroundings. 

Applying these criteria, we obtained the sample of isolated, dense groups  in real space. We place an observer in redshift space and discarded member galaxies fainter than a given flux limit, keeping those groups that have three or more members brighter than the imposed limit (z-CG$^{+3}$). 
This sample can be directly comparable to groups identified in a flux-limited mock catalogue with any observational criteria. 
We have considered the z-CG$^{+3}$ sample as the ideal sample of CGs that is expected to be recovered when identifying CGs in an observational catalogue of galaxies in redshift space.

We compared in redshift space the extracted Hickson-like compact groups (HMCG$^{+3}$, using the algorithm of \citealt{DiazGimenez+18}) to the real-space extracted dense groups (z-CG$^{+3}$).
We found that the purity and completeness of the HMCG$^{+3}$ are very low, regardless of the SAM. 
Consequently, the observational identification is strongly contaminated and fail to recover most of the ideal CGs.
The percentage of HMCG$^{+3}$ related to ideal (overdense and isolated) CGs is approximately $~ 10\%$. 
When analysing those spurious HMCG$^{+3}$ that are linked to highly-dense systems, we observed that they are indeed substructures of other loose groups, i.e, they failed to be isolated (see. Fig.~\ref{fig:bubbles} for a summary of these results). This suggests that the observational isolation criterion is quite deficient.
Furthermore, between 21 to 57\% (depending on the SAM) of the spurious HMCG$^{+3}$ are not linked to any very dense structure (regardless of the isolation). Most of them inhabit loose groups, while only $\sim 10$ per~cent are chance alignments of galaxies in filaments or the field. 

On the other hand, approximately $80$ per~cent of the ideal z-CG$^{+3}$ groups are not identified by the observational algorithm. We observe that the z-CG$^{+3}$ do not fulfil the flux constraint on the brightest galaxy or neither they have three members in a range of 3 magnitudes from the brightest, both criteria required by the observational algorithm.
The HMCG$^{+3}$ sample, meant to extract isolated dense groups, achieves this for only 10~per~cent of the CGs. Another half are dense systems embedded within looser groups, while a third are chance alignments within loose groups.
In terms of completeness, the flux limit criterion combined with the 3-magnitude range for the galaxy members prevented the identification of most of the true 3D CGs in 
redshift space.

Finally, we also showed that the purity and completeness of the automatically selected Hickson-like samples cannot be improved by changing the minimum number of members in the samples involved. The percentages of purity and completeness of HMCG$^{+4}$ relative to z-CG$^{+3}$ or z-CG$^{+4}$ are even lower than those mentioned above. 

These results suggest that, in order to increase the purity and completeness of the redshift-space samples, it seems necessary to  modify the automatic Hickson-like criteria. Nevertheless, a simple relaxation or elimination of the flux limit of the brightest galaxy and the magnitude range of the galaxy members (see for instance \citealt{Zheng+20}) could cause an increment of the completeness but at the cost of lowering the purity of the sample.  
Hence, revisiting Hickson criteria should be done with caution. If a considerable imbalance between purity and completeness cannot be avoided, then a new algorithm will be needed to achieve the main goal of maximisation of the extraction of isolation dense groups from redshift-space catalogues. These options will be explored in a future work.

\section*{Acknowledgements}
{\small  
The authors would like to thank the referee for their comments on the original manuscript.  

The Millennium Simulation databases used in this paper and the web application providing online access to them were constructed as part of the activities of the German Astrophysical Virtual Observatory (GAVO).

This work has been partially supported by Consejo Nacional de Investigaciones Cient\'\i ficas y T\'ecnicas de la Rep\'ublica Argentina (CONICET) and the Secretar\'\i a de Ciencia y Tecnolog\'\i a de la Universidad de C\'ordoba (SeCyT)}

\section*{Data Availability} 
The data underlying this article were accessed from \url{http://gavo.mpa-garching.mpg.de/Millennium/}.
The derived data generated in this research will be shared on reasonable request to the corresponding authors.

\appendix 

\bibliography{biblio}

\appendix

\section{Determination of the linking length to identify FoF-DG}
\label{app:r0}

\begin{figure}
\includegraphics[width=0.5\textwidth]{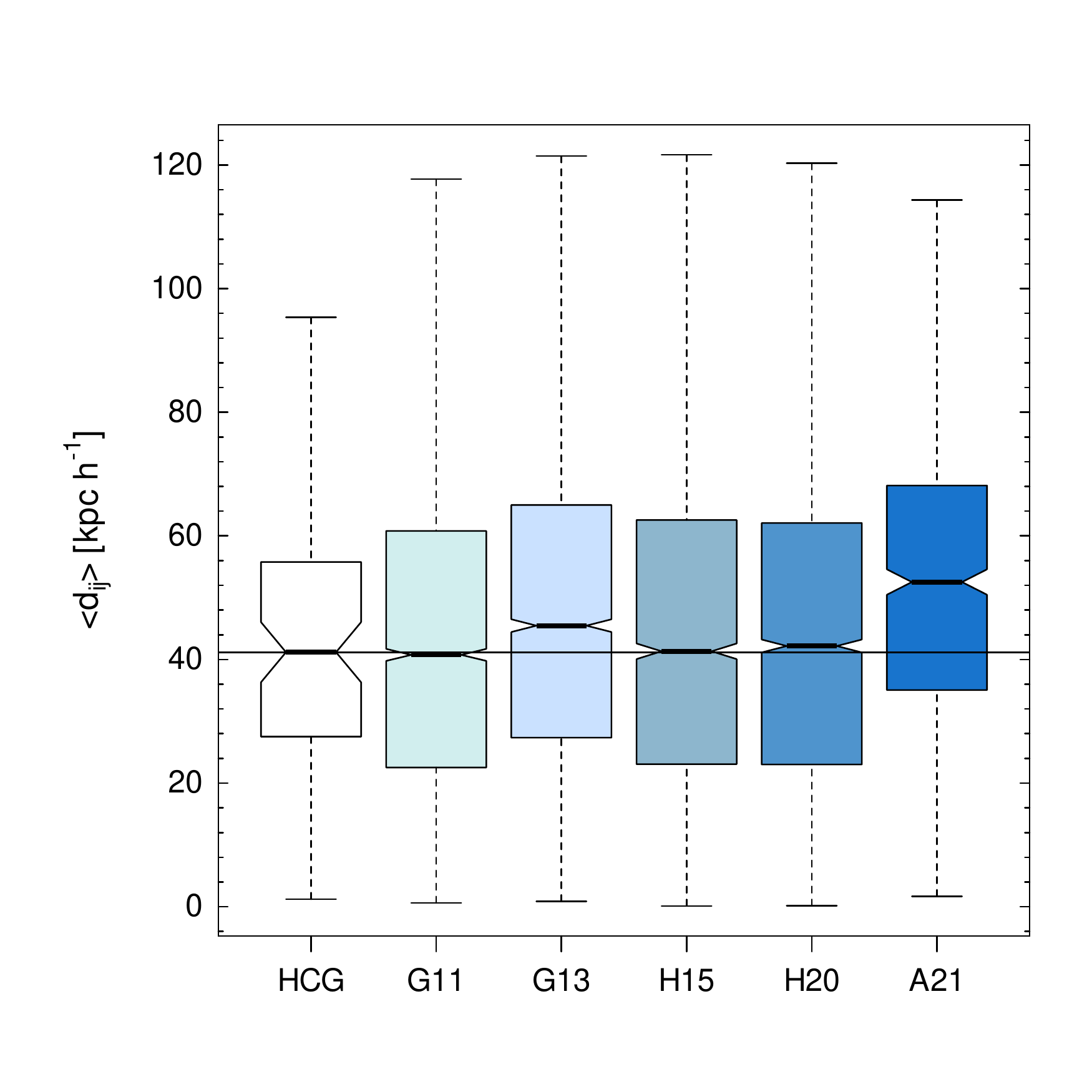} 
\caption{Median projected separation $\left\langle d_{ij}\right\rangle$ for galaxies of z-CG$^{+3}$ within the three magnitude range from the brightest,
for each SAM in comparison with Hickson CGs (HCG) 
(\emph{empty box}). The \emph{horizontal line} shows the median of the HCG sample. 
\label{fig:HCG_dij}}
\end{figure}

We intend to identify CGs in 3D that appear as compact as the visual Hickson CGs when observed in flux limited redshift space catalogues. 
Particularly, we focused on the projected separation of galaxies in HCGs. Using the well-known \cite{Hickson92} CG sample\footnote{\url{https://vizier.u-strasbg.fr/viz-bin/VizieR-3?-source=VII/213/galaxies}}, we selected a subsample of 88 HCGs that have three or more members within a range of three magnitudes from the brightest with concordant redshifts.
Following \cite{sulentic+97}, we also excluded HCG 43, 45, 46, 57, and 60 because they are not isolated. The final sample comprises 84 HCGs.
For these groups, we computed the median of the projected separation between galaxies in each group, $\left\langle d_{ij} \right\rangle$ (empty boxplot in Fig.~\ref{fig:HCG_dij}). The median of $\left\langle d_{ij} \right\rangle$ for the sample of HCGs is $\sim 41 \, h^{-1} \, {\rm kpc}$ with a $95\%$ confidence interval of [36-46]\,$h^{-1} \, {\rm kpc}$.
In this section, we describe how we use this information to find the proper value of the linking length to identify highly-dense groups in the 3D-real space with the aim of obtaining CGs that look similar to HCGs when observed in a redshift-space flux limited catalogue.

As described in section~\ref{sec:FoF-DG}, we first need to define the proper linking length $r_0$ to identify the highly-dense systems that will eventually be classified as compact groups, and when observed in redshift space their galaxy separations reproduce  the observable values obtained for the Hickson catalogue.
Therefore, we run the FoF algorithm in each simulation box for several different values of $r_0$ in the range [50-150]\,$h^{-1} \, {\rm kpc}$. 
For each sample of highly-dense groups, we applied the mass population and isolation criteria and then, we followed the procedure described in Sect.~\ref{sec:3Dmock} to construct a sample of ideal compact groups observed in a flux limited redshift space catalogue. 
These resulting samples were used to estimate the projected median separations between galaxies and were compared to the values obtained for the Hickson sample of CGs. 

From this comparison, and taking into account all the SAMs, we observed that the best resulting sample of z-CG$^{+3}$, in terms of reproducing the observational values in the best possible way, was obtained when using a linking length of $r_0=90 \,  h^{-1}\, {\rm kpc}$. 
In Fig.~\ref{fig:HCG_dij} we show the projected median separations among galaxies $\left\langle d_{ij} \right\rangle$ of z-CG$^{+3}$ for the samples obtained when using $r_0=90 \, h^{-1} \,{\rm kpc}$ in each SAM. 
For almost all the SAMS, the notches of the boxes overlap with the notches of the box of the observational sample,
indicating that the medians are in statistical agreement.
The only exception is A21, which shows a median value ($52 \, h^{-1}\, {\rm kpc}$ with a $95\%$ confidence interval of [50-55] $h^{-1} \, {\rm kpc}$), which is larger than the obtained from the observations
The medians of $\left\langle d_{ij}\right \rangle$ for rest of the SAMs vary between $41$ to $46 \, h^{-1}\, {\rm kpc}$.

\section{Comparison with previous definitions of physically dense CGs}
\label{app:reals}
\begin{table}
\setlength{\tabcolsep}{4pt}
\centering
\caption{Percentage of Real CGs in the HMCG$^{+3}$ sample.
\label{tab:reals}}
\begin{tabular}{crccc}
\hline
SAM & \multicolumn{2}{c}{HMCG$^{+3}$} && $R$ \\
\cline{2-3}
 & $N_{\rm HMCG^{+3}}$ & Reals &&  \\ 
\hline
 G11 & 10734 & 57\% &&  80\% \\ 
 G13 &  6403 & 51\% &&  72\% \\
 H15 &  5876 & 46\% &&  55\% \\
 H20 &  7711 & 51\% &&  69\% \\
 A21 &  3351 & 32\% &&  46\% \\
\hline
\end{tabular}
\parbox{\hsize}{Notes: HMCG$^{+3}$: CGs identified using the modified algorithm with Hickson criteria. 
$R = (N_{\rm R}+N_{\rm FoF-DG}^{\rm sp})/N_{\rm HMCG^{+3}}$: the percentage of HMCG$^{+3}$ that can be considered true dense systems, regardless of their 3D isolation, where $N_R$ are matching groups between zCG$^{+3}$ and HMCG$^{+3}$, and $N_{\rm FoF-DG}^{\rm sp}$ are spurious groups associated with a FoF-DG group.}
\end{table}

\begin{figure}
\includegraphics[width=0.5\textwidth]{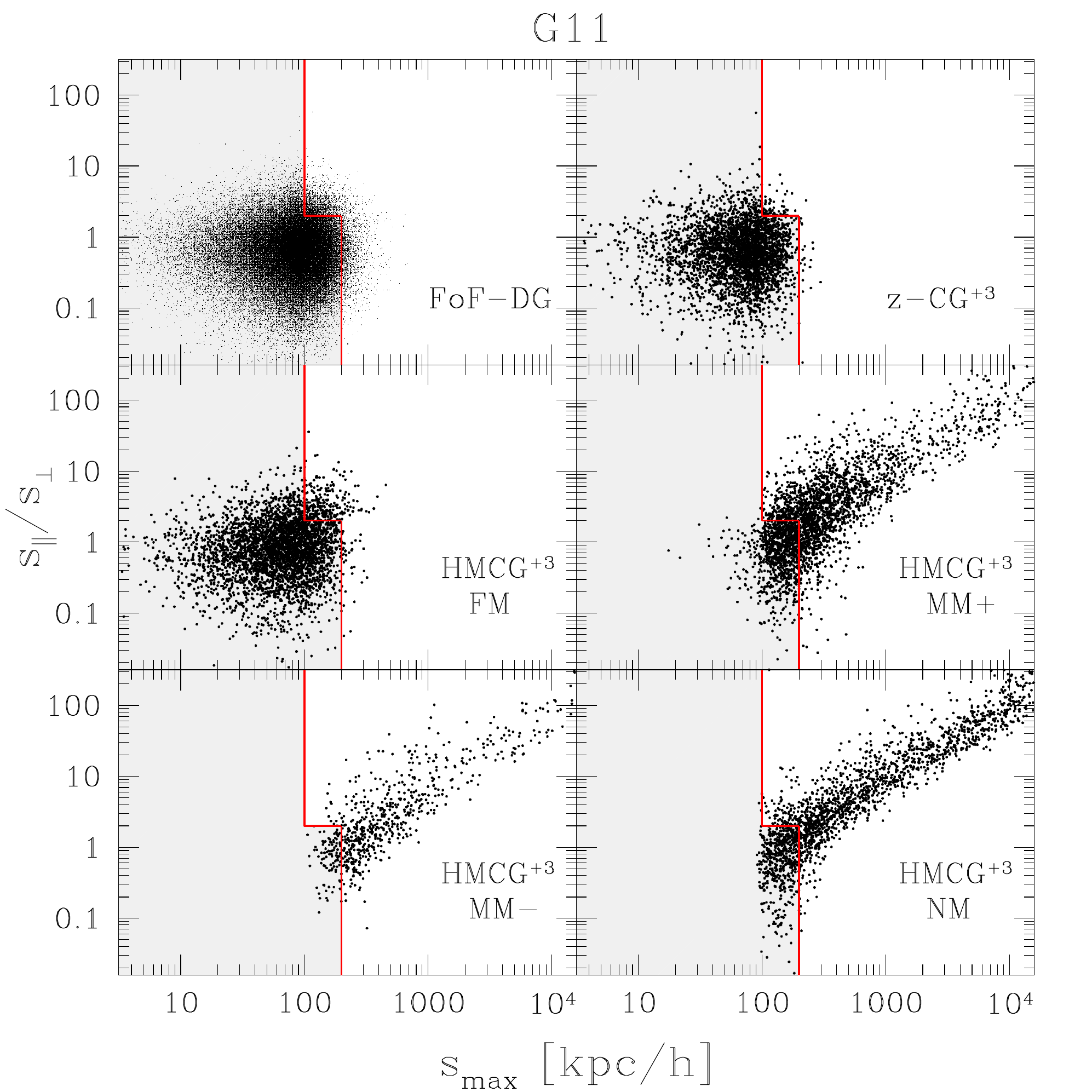}
\caption{Line-of-sight elongation versus maximum 3D interparticle separation of dense groups (FoF-DG and  z-CG$^{+3}$) and different subsamples of spurious HMCG$^{+3}$ for G11 SAM. The \emph{red line}  delimits the Real CGs (\emph{grey region}) from the chance alignments (\emph{non-shaded region}), using the classification defined by \citet{DiazGimenez&Mamon10}.}
\label{fig:sss}
\end{figure}

\cite{DiazGimenez&Mamon10} devised a criterion to split the samples of Hickson-like CGs into \emph{physically dense} \nobreak{(named Reals)} and \emph{chance alignments} (CAs) using their 3D real space information. It defined as Real CGs those whose maximum comoving 3D interparticle separation between the four closest galaxies ($s_{\rm max}$) within a magnitude range of three from the brightest galaxy is less than $100 \, h^{-1} \, \rm kpc$, or less than $200\,h^{-1}\,\rm kpc$ if the ratio of line-of-sight ($s_{\parallel}$) to transverse ($s_\perp$) sizes in real space is less than $2$. The CGs that do not satisfy the previous criterion were then classified as CAs. 

To assess the reliability of this criterion, we use the information obtained for the samples of HMCG$^{+3}$ in this work. We have extended the \cite{DiazGimenez&Mamon10} definition to systems with three galaxy members. The criterion remains the same for groups with four or more members within the 3-magnitude range; and, for triplets, it just uses the three members in the 3-magnitude range to compute the maximum comoving 3D separation and the line-of-sight and transverse sizes.

In Table~\ref{tab:reals} we quote the percentage of Real CGs in the total sample of HMCG$^{+3}$. In \citetalias{DiazGimenez+20} we found that, for HMCG with four or more members, the percentages of Real CGs vary in the range 35-56\%, while in \citetalias{DiazGimenez+21} we found that the percentages of Real CGs vary in the range 24-35\% for HMCGs with only four members (these results were for G11, G13, H15).
In this work, we obtain that the percentages of Real CGs are in the range 46-57\% for HMCGs with three or more galaxy members in all the SAMs, except for the A21 sample where the percentage lies below this range (32\%). 

Since the definition of Reals of \cite{DiazGimenez&Mamon10} only intended to classify systems by their density in real space, another way to infer the real-space density of the HMCG$^{+3}$ using the results of the present work could be analysing whether they are related to FoF-DG groups (highly dense groups in 3D). 
In the last column of Table~\ref{tab:reals} we quote the percentage of the HMCG$^{+3}$ samples that are linked with a FoF-DG group. While most of these groups do not meet the isolation criteria, this percentage can be compared with the predicted percentage of the so-called Real CGs in the previous classification. 
From this comparison, we find that the percentage of HMCG$^{+3}$ that are associated with a FoF-DG group is always higher (20\% on average) than the percentage of HMCG$^{+3}$ that were classified as Reals. Having used the association with a highly dense group as a proxy for physically dense systems, we would have predicted that between 46 to 80 per~cent of the CGs are truly dense. 
The difference between both classifications is mainly due to the fact that the matching between the HMCG$^{+3}$ with a FoF-DG is not perfect. The HMCG$^{+3}$ may include some farther interlopers as galaxy members that enlarge the value of the maximum physical distance between members, and therefore making them appear as CAs. Therefore, the classification of \cite{DiazGimenez&Mamon10} can be thought of as conservative or as a lower limit. 

But since the limiting values 
in the definition of Reals were chosen almost arbitrarily from a sample of redshift-space Hickson-like compact groups, one may wonder whether the ideal 3D CGs are indeed classified as Reals by the previous definition. 
In Fig.~\ref{fig:sss} we show scatter plots of the line-of-sight elongation versus the maximum 3D comoving separation for different groups samples in G11. 
Top panels are the FoF-DG and z-CG$^{+3}$ samples. 
From these panels, it can be seen that these very dense systems are almost perfectly  restricted to the grey region that corresponds to the location of Reals defined by \cite{DiazGimenez&Mamon10} criterion (96\% of the FoF-DG and 97\% of the z-CG$^{+3}$). 
This result is somewhat surprising because, as we mentioned before, the classification was arbitrary, however, we are showing that the highly-dense groups and our ideal (independent) z-CGs most lie within those boundaries. 

When analysing the HMCG$^{+3}$ that have been classified as spurious, a similar result is observed for the FM groups: most of these groups are Reals. 
The remaining spurious HMCG$^{+3}$ that are partially associated with a FoF-DG (MM+ and MM-) are mostly lying outside the Real's region (middle-right and bottom-left panels). As previously stated, although these systems are associated with a FoF-DG group, few galaxy interlopers produce an overestimation of 3D maximum interparticle separations, moving most of these estimates outside the Real CG (grey) zone. 

Finally, in the bottom-right panel of Fig.~\ref{fig:sss} we show the scatter plot for the HMCG$^{+3}$ classified as spurious NM groups. This plot shows that there are some NM groups that can be classified as Reals. We observe that on average $\sim 24$ per~cent of NM groups are classified as Reals (26, 27, 25, 22, and 19\% for G11, G13, H15, H20, and A21, respectively). This result implies that the classification of systems as Reals, beyond being conservative, is not free from contamination. 
Hence, it is possible to compute the percentage of HMCG$^{+3}$ classified as Reals with the previous classification (Table~\ref{tab:reals}) that are not truly dense systems: 9, 15, 24, 14, and 32\% for G11, G13, H15, H20, and A21, respectively.\\


\label{lastpage}

\end{document}